\documentclass[12pt]{iopart}
\usepackage{graphicx}

\begin{document}

\title[A model of ballistic aggregation and fragmentation]{A model of ballistic aggregation and fragmentation}

\author{Nikolay V.~Brilliantov}

\address{University of Leicester,
University Road, Leicester LE1 7RH UK and \\ Moscow State University, Vorobievy Gory 1, 119899, Moscow,
Russia} \ead{nb144@leicester.ac.uk}

\author{Anna S.~ Bodrova}

\address{Moscow State University, Vorobievy Gory 1, 119899, Moscow, Russia}
\ead{bodrova@polly.phys.msu.ru}

\author{Paul L. Krapivsky}

\address{Department of Physics, Boston University, Boston, MA 02215, USA }

\begin{abstract}
A simple model of ballistic aggregation and fragmentation is proposed. The model is  characterized by
two energy thresholds, $E_{\rm agg}$ and $E_{\rm frag}$, which demarcate different types of impacts: If
the kinetic energy of the relative motion of a colliding pair is smaller than $E_{\rm agg}$ or larger
than $E_{\rm frag}$, particles respectively merge or break; otherwise they rebound. We assume that
particles are formed from monomers which cannot split any further and that in a collision-induced
fragmentation the larger particle splits into two fragments. We start from the Boltzmann equation for
the mass-velocity distribution function and derive Smoluchowski-like equations for concentrations of
particles of different mass. We analyze these equations analytically, solve them numerically and perform
Monte Carlo simulations. When aggregation and fragmentation energy thresholds do not depend on the
masses of the colliding particles, the model becomes analytically tractable. In this case we show the
emergence of the two types of behavior: the regime of unlimited cluster growth arises when fragmentation
is (relatively) weak and the relaxation towards a steady state occurs when fragmentation prevails. In a
model with mass-dependent $E_{\rm agg}$ and $E_{\rm frag}$ the evolution with a cross-over from one of
the regimes  to another  has been detected.

\end{abstract}

\maketitle

\section{Introduction}

Collision-induced aggregation and fragmentation are ubiquitous processes underlying numerous  natural
phenomena. For a gentle collision with a small relative velocity, colliding particles can merge; a
violent collision with a large relative velocity can cause fragmentation. For intermediate relative
velocities, particles usually rebound. These collisions may still be irreversible --- the kinetic energy
could be lost in inelastic collisions. Important examples of such systems are dust agglomerates in the
Earth atmosphere or in interstellar dust clouds and proto-planetary discs
\cite{Chokshietal:1993,DominikTilens:1997,Ossenkopf:1993,SpahnetalEPL:2004}. Another example is dynamic
ephemeral bodies in planetary rings, see e.g.
\cite{Greenberg:1983,Weidenschilling:1984,Longaretti:1989}. A comprehensive description of the
aggregation and fragmentation kinetics in such systems is very complicated. Therefore it is desirable to
develop idealized models that involves three types of collisions in the simplest possible way.

The understanding of the ballistic-controlled reactions is still quite incomplete \cite{BKLR}.
Ballistic aggregation has attracted most attention (see
\cite{Ossenkopf:1993,SpahnetalEPL:2004,CPY:1990,Trizac:1995,Laurent,Laurent2,Trizac:2003,BrilliantovSpahn2006}
and references therein) and a few studies were also devoted to ballistic fragmentation (see
\cite{Red90,KrapivskyBenNaim,Pagonabarraga:2003,HidalgoPagonabarraga:2008}). The situation with
aggregation and fragmentation operating simultaneously has been analyzed only in a very special case
when all particles have the same relative velocity and the after-collision fragment mass distribution
obeys a power-law \cite{Longaretti:1989}.  Moreover, studies of pure ballistic fragmentation are usually
based on the assumption  that all particles may split independently on their mass and relative velocity
between the colliding grains \cite{KrapivskyBenNaim,Pagonabarraga:2003}. In reality, the type of an
impact strongly depends on the relative velocity \cite{SpahnetalEPL:2004}; furthermore, the agglomerates
are comprised of primary particles (``grains'') that cannot split into smaller fragments
\cite{Greenberg:1983,Weidenschilling:1984}. The fragmentation model with the splitting  probability
depending on energy has been studied in \cite{MarsiliZhang:1996}; this model, however,  does not
consider ballistic impacts  of many particles, but rather an abstract process of a successive
fragmentation of one  body with a random distribution of the bulk energy between fragments.

In this paper we propose a model of ballistic aggregation and fragmentation which accounts for three
types of collisions depending on masses and relative velocity of a colliding pair. In section 2 we
introduce the model, write the Boltzmann equation for the joint mass-velocity distribution function, and
deduce from the Boltzmann equation the rate equations for concentrations of various mass species.
Section 3 is devoted to the theoretical analysis of rate equations. Numerical verification of
theoretical results and simulation results in situations intractable theoretically is given in section
4. The last section 5 concludes the paper.

\section{The model}

Consider a system comprised  of primary particles (monomers) of mass $m_1$ and radius $r_1$, which
aggregate to form clusters of $2,\, 3, \ldots, k, \ldots $ monomers with masses $m_k=km_1$. In some
applications (e.g. in modeling of dynamic ephemeral bodies \cite{Greenberg:1983,Weidenschilling:1984})
it is appropriate to consider clusters as objects with fractal dimension $D$; for compact clusters
$D=3$.  The characteristic radius of an agglomerate containing $k$ monomers scales with mass as $r_k
\sim r_1 k^{1/D}$. We assume that when the kinetic energy of two colliding clusters in the
center-of-mass reference frame (the ``relative kinetic energy'' in short) is less than $E_{\rm agg}$,
they merge. In this case a particle of mass $(i+j)m_1$ is formed. If the relative kinetic energy is
larger than $E_{\rm agg}$, but smaller than $E_{\rm frag}$, the colliding particles rebound without any
change of their properties. Finally, if the relative kinetic energy exceeds $E_{\rm frag}$, {\it one} of
the particles  (we assume that the {\it larger one}) splits into two fragments.
We denote $p_{i,k-i}$ the probability that a particle of mass $k$ splits into particles of masses  $i$
and $k-i$. Obviously, $\sum_i p_{i,k-i}=1$ and $p_{i,k-i}=0$ if $k\le i$.

We restrict ourselves to dilute and spatially uniform systems. Let $f_i \equiv f(\vec{v}_i,t)$ be the
mass-velocity distribution function which gives the concentration of particles  of mass $m_i$ with the
velocity $\vec{v}_i$ at time $t$. The mass-velocity distribution function evolves according to the
Boltzmann equation
\begin{equation}
\label{eq:BEgen} \frac{\partial}{\partial t} f_k\left(\vec{v}_k, t \right) = I^{\rm agg}_{k}+I^{\rm reb}_{k}+I^{\rm frag}_{k}\, ,
\end{equation}
where $I^{\rm agg}_{k}$, $I^{\rm reb}_{k}$ and $I^{\rm frag}_{k}$ are respectively the collision
integrals  describing collisions leading to aggregation, rebound, and fragmentation. The first integral
reads
\begin{eqnarray}
\label{eq:Iagg}
I^{\rm agg}_{k}(\vec{v}_k)  &=& \frac12 \sum_{i+j=k}  \sigma_{ij}^2 \int d
\vec{v}_i\int d \vec{v}_j \int d\vec{e} \, \Theta\left(-\vec{v}_{ij} \cdot \vec{e} \, \right)
\left|\vec{v}_{ij} \cdot \vec{e} \,\right| \times \\    & \times & f_i\left(\vec{v}_i \right)
f_j\left(\vec{v}_j \right)\Theta\left(E_{\rm agg}-E_{ij} \right) \delta(m_k \vec{v}_k-m_i \vec{v}_i-m_j
\vec{v}_j )
\nonumber \\
&-& \sum_{j}\sigma_{kj}^2 \int d \vec{v}_j \int d\vec{e} \, \Theta\left(-\vec{v}_{kj} \cdot \vec{e} \,
\right) \left|\vec{v}_{kj} \cdot \vec{e} \,\right|  \times    \nonumber \\    &\times&
f_k\left(\vec{v}_k
\right) f_j\left(\vec{v}_j \right)
   \Theta\left(E_{\rm agg}-E_{kj}  \right) \nonumber   \, .
\end{eqnarray}
Here $\sigma_{ij}=r_1\left( i^{1/D}+j^{1/D} \right)$ is the sum of radii of the two clusters,  while
$m_k=m_i+m_j$ and  $m_k \vec{v}_k =\vec{v}_im_i +m_j \vec{v}_j$, due to the conservation of mass and
momentum. We also introduce the relative velocity, $\vec{v}_{ij}=\vec{v}_i-\vec{v}_j$, the reduced mass,
$\mu_{ij}=m_im_j/(m_i+m_j)$, and the relative kinetic energy, $E_{ij} =\frac12 \mu_{ij}v_{ij}^2$. The
unit vector $\vec{e}$ specifies the direction of the inter-center vector at the collision instant. The
factors in the integrand in Eq.~(\ref{eq:Iagg}) have their  usual meaning (see e.g.
\cite{BrilliantovPoeschelOUP}): $\sigma_{ij}^2 \left|\vec{v}_{ij} \cdot \vec{e} \,\right|$ defines the
volume of the collision cylinder, $\Theta\left(-\vec{v}_{ij} \cdot \vec{e} \, \right)$ selects only
approaching particles and $\Theta\left(E_{\rm agg}-E_{ij}  \right)$ guarantees that the relative kinetic
energy does not exceed $E_{\rm agg}$ to cause the aggregation. The first sum in the right-hand side of
Eq.~(\ref{eq:Iagg}) refers to collisions where a cluster of mass $k$ is formed from smaller clusters of
masses $i$ and $j$, while the second sum describes the collisions of $k$-clusters with all other
aggregates.

For collisions leading to fragmentation we have
\begin{eqnarray}
\label{eq:Ifrag}
I^{\rm frag}_{k} (\vec{v}_k)& =&\sum_{j}\sum_{i \le j}p_{k,j-k}\left(1- \frac12
\delta_{i,j}\right) \sigma_{ij}^2 \int d \vec{v}_j\int \!d \vec{v}_i \int  d \vec{e} \,
\Theta\left(-\vec{v}_{ij} \cdot \vec{e} \, \right)  \times
   \nonumber \\
   &\times& \left|\vec{v}_{ij} \cdot \vec{e} \,\right| f_j\left(\vec{v}_j\right)
   f_i\left(\vec{v}_i \right)
   \Theta\left(E_{ij} - E_{\rm frag} \right)\Delta (\vec{v}_i,\, \vec{v}_j,  \, \vec{v}_k) \\
   &-&\sum_{i\le k}\left(1- \frac12 \delta_{i,k}\right)\sigma_{ki}^2 \int d \vec{v}_i \int d\vec{e} \,
   \Theta\left(-\vec{v}_{ki} \cdot \vec{e} \, \right) \left|\vec{v}_{ki} \cdot \vec{e} \,\right| \times
   \nonumber \\
   &\times& f_k\left(\vec{v}_k\right)
   f_i\left(\vec{v}_i \right)
   \Theta\left(E_{ki} - E_{\rm frag} \right)  \nonumber \, ,
\end{eqnarray}
where $m_j =m_k+m_{j-k}$ and we use the abbreviation, $\Delta(\vec{v}_i,\, \vec{v}_j,  \,
\vec{v}_k)=\delta(m_j \vec{v}_j+ m_i\vec{v}_i - m_k \vec{v}_k- m_{k-j}\vec{v}_{k-j}^{\,\prime} +
m_i\vec{v}_i^{\,\prime})$  for the factor which guarantees the momentum conservation at the collision.
The after-collisional velocities $\vec{v}_{k-j}^{\,\prime}$ and $ \vec{v}_i^{\,\prime}$ are determined
by a particular fragmentation model.  The first sum in Eq.~(\ref{eq:Ifrag}) describes the collision of
particles of mass $i$ and  $j$ ($j \ge k$, $j\ge i$) with the relative kinetic energy above the
fragmentation threshold $E_{\rm frag}$. The larger particle, i.e. the particle of mass $j$, splits with
the probability $p_{k, j-k}$ into two particles of mass $k$ and $j-k$ thereby giving rise to a particle
of mass $k$. The second sum describes the opposite process, when particles of mass $k$ break in
collisions with smaller particles. In the present study we do not need an explicit expression for the
velocities $\vec{v}_{k-j}^{\,\prime}$ and $ \vec{v}_i^{\,\prime}$ of the fragments. We do not also need
an expression for the collision integral $I^{\rm reb}_{k}$; it has the usual form (see e.g.
\cite{BrilliantovPoeschelOUP}) with a slight modification to account for the requirement that the
relative kinetic energy $E_{ij}$ belongs to the interval  $(E_{\rm agg} \, < E_{ij} < \, E_{\rm frag})$.

Thus we have a mixture of particles of different masses and each species generally has its own
temperature.  For this (granular) mixture we write
\begin{equation}
\label{eq:nTiTdef}
 n_i  =  \int d\vec{v}_i f_i (\vec{v}_i) \, ,  \qquad \qquad
        ~~N=\sum_i n_i
\end{equation}
where $n_i$ is the number density (concentration) of particles of mass $i$ and $N$ is the total number
density. Using the mean kinetic energy of different species one can also define the partial granular
temperatures $T_i$ for clusters  of mass $i$  and effective temperature $T$ of the mixture
\cite{GarzoDufty:1999}.  We assume that the distribution function $f_i(\vec{v}_i,t )$ may be written as
\cite{SpahnetalEPL:2004,BrilliantovSpahn2006,GarzoDufty:1999}
\begin{equation}
\label{eq:NDFpartial} f_i(\vec{v}_i, t) = \frac{n_i(t)}{v_{0, \, i}^3(t)} \phi_i (\vec{c}_i) \,, \qquad
\vec{c}_i \equiv \frac{\vec{v}_i}{v_{0,\,i}} \, ,
\end{equation}
where $v_{0, \,  i}^2(t) = 2T_i(t)/m_i$ is the thermal velocity and $ \phi (c_i) $ the reduced
distribution  function. For the force-free granular mixtures the velocity distribution functions of the
components are not far from the Maxwellian distribution  \cite{GarzoDufty:1999}, which reads in terms of
the reduced velocity $\vec{c}=\vec{v}/v_T$,
\begin{equation}
\label{eq:Maxwel} \phi_M(\vec{c}) = \pi^{-3/2}\exp(-c^2)\, .
\end{equation}
The equipartition between different components may, however, break down, in the sense that the partial
temperatures $T_i$ are not equal and  differ from the effective temperature $T$ \cite{GarzoDufty:1999}.
Here we ignore the deviation from the Maxwellian distribution \footnote{Note that for a slightly
modified model, where only in a small fraction of collisions that fulfil the aggregation criterion,
particles merge and only in a small fraction  of collisions that fulfil the fragmentation criterion,
particles split, the velocity distribution is close to the Maxwellian, like  in a granular mixture. At
the same time this modified model would lead to the same kinetic equations (9), but with the
renormalized time scale.} and possible violation of the equipartition and use the approximation, $\phi_i
(c_i) \approx \phi_M(c_i)$ and $T_i \approx T$ for all $i$. We also assume that the temperature of the
system does not depend on time. This is formally inconsistent within the Boltzmann equation
(\ref{eq:BEgen}), yet in many applications the temperature is approximately constant on a astronomical
time scale, e.g. this happens in planetary rings where the viscous heating due to  the shearing mode of
the particle orbital motion keeps the granular temperature constant
\cite{PlanetaryRings:1984,SpahnetalEPL:2004}. The consistent approach would be to modify the Boltzmann
equation to take into account gradients of the local hydrodynamic velocity, which will result in
additional terms in the velocity distributions $f_i$, proportional to these gradients. If we assume that
such  gradients are very small but still sufficient to support constant temperature due to viscous
heating, we can neglect the small corrections to the distribution functions and approximate them with a
gradient-free form (\ref{eq:NDFpartial}).

Integrating Eq.~(\ref{eq:BEgen}) over $\vec{v}_k$ we obtain the equations for the zero-order moments of
the  velocity distribution functions $f_k$, that is, for the concentrations $n_k$. Taking into account
that collisions resulting in rebounds do not change the concentrations of different species and using
(\ref{eq:Iagg})--(\ref{eq:Ifrag}) together with (\ref{eq:nTiTdef})--(\ref{eq:Maxwel}) we arrive at rate
equations
\begin{eqnarray}
\label{eq:Smoluch}
\frac{d}{d t} n_k &=&  \frac12  \sum_{i+j=k}C_{i,j} n_in_{j} -n_k \sum_{i=1}^{\infty}
C_{k,i} n_i +\sum_{j=k+1}^{\infty}\!\sum_{i=1}^j A_{i,j}n_i n_j\left(1- \!\frac12
\delta_{i,j}\right) p_{k,j-k} \nonumber \\
&-&n_k (1-\delta_{1k}) \sum_{i=1}^{k} A_{k,i} n_i \left(1- \frac12 \delta_{i,k}\right) \, ,
\end{eqnarray}
with rates given by
\begin{eqnarray}
  \label{eq:KinCoef}
&&C_{i,j}=2\sigma_{ij}^2 \left( \frac{2T\pi}{\mu_{ij}} \right )^{1/2} \left( 1- \left(1+\frac{E_{\rm
agg}}{T} \right) e^{-E_{\rm agg}/T} \right)
\nonumber \\
&&A_{i,j}= 2\sigma_{ij}^2 \left( \frac{2T\pi}{\mu_{ij}} \right )^{1/2} e^{-E_{\rm frag}/T}  \,.
\end{eqnarray}
 It is useful to verify that the above kinetic equation (\ref{eq:Smoluch}) fulfills the condition
of mass conservation, $\sum_k km_1n_k =M={\rm const.}$, where $M$ is the total mass density.

The probability of splitting $p_{ik}$ depends on geometric and mechanical properties of the aggregates
and  generally it is quite complicated.  For concreteness we focus on splitting into (almost) equal
fragments. Namely, we assume that a particle of mass $2km_1$  splits into two equal halves, while a
particle of mass $(2k+1)m_1$ splits into particles of mass $km_1$ and $(k+1)m_1$. For this  choice of
the splitting probability, the kinetic equation reads
\begin{eqnarray}
\label{eq:Smoluch2}
\frac{d}{d t} n_k &=&  \frac12  \sum_{i+j=k}C_{i,j} n_in_j -n_k \sum_{i=1}^{\infty}
C_{k,i} n_i -n_k \sum_{i=1}^{k} A_{k,i} n_i \left(1- \delta_{k,i}/2\right)
 \nonumber \\
&+&2 \sum_{i=1}^{2k} A_{2k,i} n_{2k}n_i \left(1- \delta_{2k,i}/2\right) + \sum_{i=1}^{2k+1}
A_{2k+1,i} n_{2k+1}n_i \left(1- \delta_{2k+1,i}/2\right)  \nonumber \\
&+& \sum_{i=1}^{2k-1} A_{2k-1,i} n_{2k-1}n_i \left(1- \delta_{2k-1,i}/2\right) \, .
\end{eqnarray}
In the next sections we study this model theoretically  and  numerically.

\section{Theoretical analysis}

To understand the qualitative behavior it is instructive to start with the simplest
model which allows an analytical treatment.

\subsection{Constant rates}

Consider first the model with constant rates $A_{i,j}$ and $C_{i,j}$. Without loss of
generality we can choose these rates to be
\begin{equation}
\label{eq:Kineq_simple} C_{i,j}= 2  \, , \qquad \qquad A_{i,j}= 2 \lambda \, ,
\end{equation}
The parameter $\lambda$ quantifies the relative intensity of fragmentation with respect to aggregation.
Fragmentation prevails when $\lambda>1 $ while aggregation wins in the opposite case of $\lambda<1 $. If
$\lambda=1 $ two processes are in a balance.

Even in this simple case we still ought to analyze a cumbersome system of infinitely many equations. To
gain insight it is useful to consider the evolution of the total density $N=\sum n_k$. (In many problems
involving aggregation and fragmentation this quantity satisfies a simple equation that does not contain
other densities.)  Summing up all equations (\ref{eq:Smoluch2})  we obtain
\begin{equation}
\label{eq:dNdt}
\dot{N} =  -(1-\lambda)N^2 -  \lambda n_1^2
\end{equation}
which has indeed a neat form, although it additionally involves the density of monomers. This density
evolves  according to
\begin{equation}
\label{eq:dn1dt} \dot{n}_1 = -2n_1 N + 2 \lambda \left[n_2(2n_1+n_2) +n_3(n_1+n_2)+\frac12 n_3^2 \,
\right] \, .
\end{equation}
Although we do not have a closed system we can already reach some qualitative conclusions.  Equation
(\ref{eq:dNdt})  indicates that two regimes are possible. If $\lambda < 1$, i.e. when aggregation
prevails, the system continues to evolve leading to  formation of larger and larger clusters; when
$\lambda > 1$, one expects that the system reaches a steady state. We now analyze these situations in
more detail.

\subsubsection{Unlimited cluster growth, $\lambda <1$}

In this case larger and larger clusters will arise. Since the total mass is conserved, one expects that
the  concentration of small clusters will rapidly decrease. Therefore,  $n_1\ll N$ when $t \gg 1$ and
therefore one can omit the second term on the right-hand side of (\ref{eq:dNdt}). Similarly one can keep
only the first term on the right-hand side of (\ref{eq:dn1dt}). This leads to the simplified equations
\begin{equation}
\label{eq:dNdt1}
\dot{N} \simeq -(1-\lambda)N^2,  \qquad  \qquad\dot{n}_1 \simeq -2n_1 N
\end{equation}
which are solved to yield the large time behavior:
\begin{eqnarray}
\label{eq:Nt}
N \simeq  \frac{1}{(1-\lambda) t } \\
\label{eq:n1t}
 n_1 \sim t^{-2/(1-\lambda)}
\end{eqnarray}

Further, one anticipates that the density distribution approaches the scaling form
\begin{equation}
\label{eq:scal_nk}
n_k \simeq t^{-2z}\Phi(x), \quad x = \frac{k}{t^z}
\end{equation}
in the scaling limit $t \to \infty$, $k\to \infty$, with the scaled mass $x=k/t^z$ kept finite.
(Here $z$ is the dynamic exponent characterizing the average mass:
$\langle k\rangle \sim t^z$.) The scaling form agrees with mass conservation:
$\sum kn_k\simeq \int dx\,x\Phi(x)$ is manifestly time-independent.

The exponent $z$ can be found from the known asymptotic behavior of $N(t)$. Indeed, writing
\begin{equation*}
N = \sum_{k\geq 1}n_k \simeq t^{-z} \int_0^{\infty} dx\,\Phi(x) \sim t^{-z}
\end{equation*}
and matching this with already known asymptotic behavior (\ref{eq:Nt}) we conclude that  $z=1$. If we
further assume that $\Phi(x) \sim x^{\mu}$ for $x \ll 1$ and combine this asymptotic with $z=1$ and the
scaling ansatz  (\ref{eq:scal_nk}) we obtain $n_1 \sim t^{-2} \, t^{-\mu}$. Matching with
(\ref{eq:n1t}) we get $\mu =2\lambda/(1-\lambda)$. Therefore
\begin{equation}
\label{eq:n_k} n_k \sim \frac{1}{t^2}\, \left( \frac{k}{t}  \right)^{2\lambda/(1-\lambda)}
\end{equation}
when $k\ll t$. Obviously, the above equation implies the asymptotic time dependence $n_k \sim
t^{-2/(1-\lambda)}$  and the mass dependence $n_k \sim k^{2\lambda/(1-\lambda)}$ for  $x=k/t \ll 1$.

\subsubsection{Relaxation to a steady state, $\lambda >1$}

For $\lambda >1$ the system evolves to a steady state with constant concentration of clusters.  In this
case $\dot{n}_k=\dot{N}=0$ and Eq. (\ref {eq:dNdt}) yields,
\begin{equation}
\label{eq:N_n}
n_1 = N \sqrt{1-\lambda^{ -1}}
\end{equation}
The densities $n_k$ rapidly decay with $k$. Therefore $n_{2k} \ll n_k$ for $k \gg 1$ and the governing
equations (\ref{eq:Smoluch2}) for the stationary concentrations simplify to
\begin{equation}
\label{eq:stationar}
\sum_{i=1}^{k-1}n_in_{k-i} -2(1+\lambda) n_k N = 0
\end{equation}
where we have approximated a finite sum up to $k \gg 1 $ by an infinite sum and ignore the terms
containing $n_{2k}, n_{2k \pm 1}$. The above equation is supposed to be valid for large $k$; it is
certainly invalid for $k=1$ when the right-hand side does not vanish. The qualitative form of the large
$k$ asymptotic behavior is determined by the mathematical structure of  (\ref{eq:stationar}). To extract
this asymptotic let us consider the simplest version when  Eq. (\ref{eq:stationar}) is valid for all
$k\geq 2$. Specifically, let us probe the model
\begin{equation}
\label{eq:close_model1}
   \sum_{i=1}^{k-1}n_in_{k-i} -2(1+\lambda) n_k N =-(1+2\lambda)N\delta_{k,1} \, ,
\end{equation}
where the amplitude $(1+2\lambda)$ was chosen to set $N=1$. (For model (\ref{eq:close_model1}),  this
choice merely sets the overall amplitude.)

The infinite system (\ref{eq:close_model1}) forms a recurrence and therefore it is solvable.
Introducing the generating function
\begin{equation}
\label{eq:Nz}
\mathcal{N}(z)=\sum_{k\geq 1} n_k\,z^k
\end{equation}
we recast (\ref{eq:close_model1}) into a quadratic equation
\begin{equation}
\label{eq:Nz_z}
\mathcal{N}^2-2(1+\lambda)\mathcal{N} + (1+2\lambda)z=0
\end{equation}
which is solved to yield
\begin{equation}
\label{eq:Nz_solution}
\mathcal{N} = (1+\lambda)- \sqrt{(1+\lambda)^2 -(1+2\lambda)z}
\end{equation}
Expanding ${\cal N}(z)$ we arrive at
\begin{equation}
\label{eq:nk_sol} n_k=\frac{1+\lambda}{\sqrt{4\pi}}
\left[1-\frac{1}{(1+\lambda)^2}\right]^k\, \frac{\Gamma(k-\frac{1}{2})}{\Gamma(k+1)}
\end{equation}
From this solution one gets $n_1=(\lambda +1/2)/(\lambda +1)$,  which of course directly follows  from
Eq.~(\ref{eq:close_model1}) as well.  For large $k$, equation (\ref{eq:nk_sol}) simplifies to
\begin{equation}
\label{eq:nk_sol_asymp} n_k \simeq \frac{1+\lambda }{\sqrt{4\pi}}\,k^{-3/2}
\left[1-\frac{1}{(1+\lambda)^2}\right]^k
\end{equation}

We considered other tractable versions when Eq.~(\ref{eq:stationar}) is exact above a certain
threshold, $k\geq \kappa+1$, while for $k=1,\ldots,\kappa$ we use the same modification as in
Eq.~(\ref{eq:close_model1}) for $k=1$. In this case instead of (\ref{eq:Nz_z}) one gets
$\mathcal{N}^2-2(1+\lambda)\mathcal{N} + P(z)=0$ with $P(z)=A_1z+\ldots+A_\kappa z^\kappa$. The root of
$P(z)=(1+\lambda)^2$ closest to the origin is positive [one can show that it exceeds unity, $z_*>1$]
and non-degenerate. Expanding the generating function $\mathcal{N} = 1+\lambda- \sqrt{(1+\lambda)^2
-P(z)} $ leads to the asymptotic $n_k\sim k^{-3/2} z_*^{-k}$. The above argument favors the asymptotic
behavior
\begin{equation}
\label{eq:nk_asymp} n_k \simeq Ak^{-3/2}e^{-\gamma k}
\end{equation}
This asymptotic form is {\em universal} and only the parameters $A,\, \gamma$ depend on the specificity
of the model, that is on the parameter $\lambda$.

It is impossible to determine $A,\, \gamma$ since models like  (\ref{eq:close_model1}) are uncontrolled
approximations. Let us still use such models and choose the simplest one which obeys the exact relation
of Eq.~(\ref{eq:N_n}). The model (\ref{eq:close_model1}) is inappropriate as it fails to satisfy
(\ref{eq:N_n}): $(\lambda +1/2)/(\lambda +1)>\sqrt{1-\lambda^{-1}}$. Modifying Eq.~(\ref{eq:stationar})
at $k=1,2$ yields (we still set $N=1$)
\begin{equation}
\label{eq:B}
\sum_{i+j=k} n_i n_j  -2(1+\lambda)n_k = - q\delta_{k,1} - (2\lambda +1 -q)\delta_{k,2}
\end{equation}
with
\begin{equation*}
q = 2(1+\lambda)\sqrt{1-\lambda^{-1}}
\end{equation*}
ensuring the validity of  (\ref{eq:N_n}).
The same approach as before gives (\ref{eq:nk_asymp}) with  $A$ and $\gamma$. In particular,
\begin{equation}
\label{eq:aB}
\gamma = \ln \left[(1+\lambda)\,
\frac{\sqrt{1-\lambda^{-1}+\Lambda} - \sqrt{1-\lambda^{-1}}}{\Lambda}\right]
\end{equation}
where we have used the short-hand notation
$\Lambda = 1+2\lambda - 2(1+\lambda)\sqrt{1-\lambda^{-1}}$.
For $\lambda=2$ (which we have studied numerically) one gets
\begin{equation}
\label{eq:gamma}
 \gamma= 0.495156\ldots
\end{equation}
This is an uncontrolled approximation, of course. Interestingly, the result is rather close to the
numerically obtained value $\gamma\approx 0.465$.

\subsection{Mass-independent energy thresholds}

We now turn to the analysis of the situation when aggregation and fragmentation energy  thresholds
$E_{\rm agg}$ and $E_{\rm frag}$ are constant. In this case the total density of clusters  evolves
according to
\begin{equation} \label{eq:N_n_ij}
 \dot{N} = -\frac12 (1-\lambda) \sum_{i=1}^{\infty}\sum_{j=1}^{\infty}
  C_{i,j}n_i n_j  - \frac12 \lambda C_{1,1}n_1^2 \,  ,
\end{equation}
where $\lambda^{-1} = e^{E_{\rm frag}/T} \left(1-(1+E_{\rm agg}/T \right)e^{-E_{\rm agg}/T})$. This
again implies the existence of the two opposite evolution regimes: For $\lambda >1$ the relaxation to a
steady state is expected, while for $\lambda <1$ -- the regime of the unlimited cluster growth.

The rates $C_{ij}=C(i,j)$ and $A_{ij}=A(i,j)$ differ by a constant factor $\lambda$;
moreover, they are homogeneous functions of their arguments:
\begin{equation}
\label{eq:Scal_AC} A(a i, a j) = a^{\nu} A(i,j) \, ,  \qquad \qquad C(a i, a  j) = a^{\nu} C(i,j)
\end{equation}
with the exponent
\begin{equation}
\label{eq:nu} \nu = \frac{2}{D} -\frac12 \, ,
\end{equation}
which  follows from the relation for a particles mass $m_i = im_1$, the cross-section of the collision
cylinder, $\sigma_{ij}^2 \sim \left(i^{1/D}+ i^{1/D} \right)^2$, and Eqs.~(\ref{eq:KinCoef}). Plugging
the scaling Ansatz (\ref{eq:scal_nk}) into Eq.~(\ref{eq:Smoluch2}), taking into account that for $k \gg
1$ the summation may be approximated by integration, and exploiting the homogeneity of the rate kernels,
Eq.~(\ref{eq:Scal_AC}), we obtain (see e.g. \cite{vanDongenErnst:1985,ChengRedner:1988,Leyvraz:2003} for
analysis of similar integro-differential equations for the re-scaled mass distribution)
\begin{eqnarray}
\label{eq:Phi} &&\frac{z}{t^{2z+1}} \left( 2\Phi(x) +  x \Phi^{\,\prime}(x) \right) =
\frac{1}{t^{(3-\nu)
z}} \, \int_0^{\infty} dy \Phi (y) \left[ (C_{x,y}+A_{x,y}) \Phi(x) \right.\\
&&~~~~~~~~~~~~~~~~~~~~~~~~~~~~~~~~~~~~~~\left. -\frac12 C_{y,x-y}\Phi(x-y) -4 A_{2x,y} \Phi(2 x) \right]
\nonumber \, .
\end{eqnarray}
{}From Eq. (\ref{eq:Phi}) we find the scaling exponent
\begin{equation}
\label{eq:z1} z=\frac{1}{1-\nu} = \left( \frac32 -\frac2D  \right)^{-1}  \, .
\end{equation}
Although the scaling theory does not allow to determine the scaling function $\Phi(x)$, one can find the
total concentration of clusters from Eq. (\ref{eq:z1}):
\begin{equation}
\label{eq:N_z_sc} N(t) \sim t^{-z} \sim t^{-2D/(3D-4)} \, .
\end{equation}
Correspondingly, the average clusters mass grows as $\left< k \right> = M/N \sim t^z$. 

\subsection{Dependence of the energy thresholds on masses of colliding particles}

In the preceding analysis we have assumed that $E_{\rm agg}$ and $E_{\rm frag}$ do not depend on the
mass of colliding particles so that $\lambda_{ij}= A_{ij}/C_{ij} =\lambda$ is constant. In reality,
however, such dependence does exist,  implying that $\lambda_{ij}$ is a function of $i$ and $j$. Still,
if $\lambda_{ij} >1$ or $\lambda_{ij}<1$ for  all $i$ and $j$, the qualitative behavior of a system is
similar to that for the case of constant $E_{\rm agg}$ and $E_{\rm frag}$: For $\lambda_{ij} >1$ a
relaxation to a steady state is expected, while for $\lambda_{ij}<1$ an unlimited cluster growth is
observed. The most interesting behavior is expected when $\lambda_{ij}-1$ changes its sign with
increasing clusters masses $i$ and $j$. In this case one anticipates a cross-over from one type of
evolution to another.

To choose a realistic dependence of $E_{\rm agg}$ and $E_{\rm frag}$
 on the masses of colliding particles, one needs more details of the collision process. We shall use
 the threshold energy for ballistic aggregation that takes into account  surface  adhesion \cite{BrilAlbSpaPoes:2007}.
 In this case
\begin{equation}
 \label{eq:E_agg_ij}
E_{\rm agg}(i,j)=E_{\rm agg}^{0}\left(\frac{ij}{i+j}\right)^\frac{4}{3} \, ,
\end{equation}
where $E_{\rm agg}^{0}$ is expressed in terms of the monomer radius, particle surface tension, the Young
modulus and the Poisson ratio of the particle material
(see \cite{BrilAlbSpaPoes:2007} for the explicit expression for $E_{\rm agg}^{0}$).

For the energy of fragmentation we assume that it is equal to the energy required to create an
additional surface, which may be roughly estimated as twice the  area of the equatorial cross-section of
the larger particle (recall that the model assumes, that the larger particle in a collision pair breaks
down). Hence we adopt the following mass dependence for $E_{\rm frag}$:
\begin{equation}
\label{eq:E_frag_ij} E_{\rm frag}(ij)=E_{\rm frag}^{0}\left(\theta_{ij}i + \theta_{ji} j \right)^2 \, ,
\end{equation}
where $\theta_{ij}=1$ if $i>j$,  $\theta_{ij}=0$ if $i<j$ and $\theta_{ii}=1/2$; $E_{\rm
frag}^{0}=2\pi\gamma_s r_1^2$, with $\gamma_s$ being the surface tension.

\section{Numerical simulations}

In our numerical studies we apply two different approaches -- the solutions of the system of
differential equations and the direct modeling of random  aggregation and fragmentation processes (with
the corresponding rates $C_{i,j}$ and $A_{i,j}$) by means of Monte Carlo (MC) method. In the former case
we use $1000$ equations and in the later one $100,000$ monomers (we always used the mono-disperse
initial conditions). The approach based on the solution of differential equation has an obvious
deficiency as one must approximate an infinite system of equations with a finite one. The MC approach is
more time consuming, yet it has an advantage of directly imitating the physical processes in which
particles are involved. To model the fragmentation and aggregation kinetics by MC we use the standard
Gillespie algorithm \cite{Gillespie:1976,Feistel:1977} (see \cite{PoeschelBrilliantovFrommel:2003} for
the application of this algorithm to the aggregation and fragmentation processes).

\begin{figure}[htbp]
\centerline{
    \includegraphics[width=0.65\columnwidth]{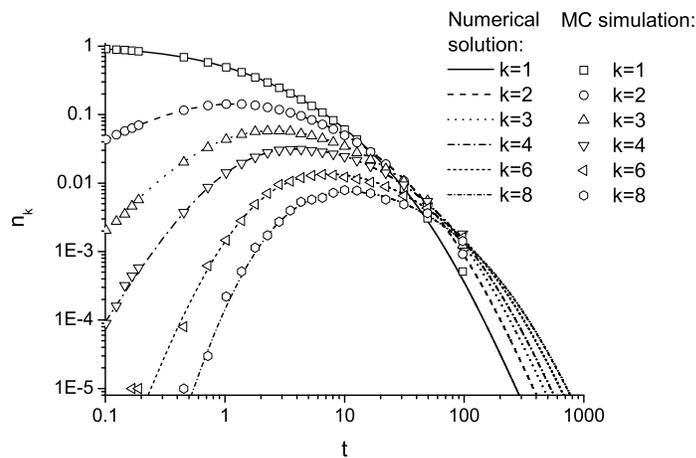}
    }
    \caption{
    Evolution of cluster concentrations $n_k$ in the case of constant rates
    with $\lambda =0.5$. The lines correspond to the numerical solution of 1000 differential equations,
    symbols are the results of  MC simulation with 100 000 monomers for the monodisperse initial
    conditions. Note, that while $n_1(t)$ always decays, $n_k(t)$ initially increase and than decay to
    zero.
    }
  \label{C1A05MC}
\end{figure}

\begin{figure}[htbp]
  \centerline{
    \includegraphics[width=0.65\columnwidth]{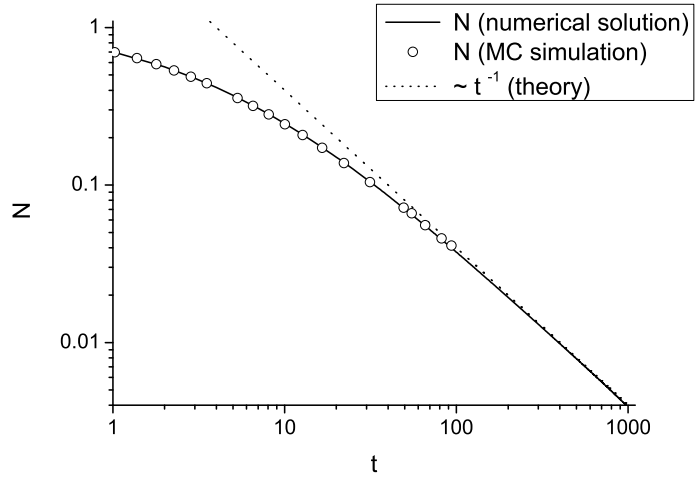}
    }
    \caption
    {Evolution of the  total number of clusters $N$ for the same system as in Fig.~\ref{C1A05MC}.
    The solid line corresponds to the numerical solution of the differential equations,
    symbols  -- to the MC simulation and  the dotted line shows the theoretical prediction, Eq.\, (\ref{eq:Nt}),
    $N(t)\sim  t^{-1}$ for $t \gg 1$.
    }
  \label{C1A05Nkl}
\end{figure}

\begin{figure}[htbp] 
  \centerline{
    \includegraphics[width=0.65\columnwidth]{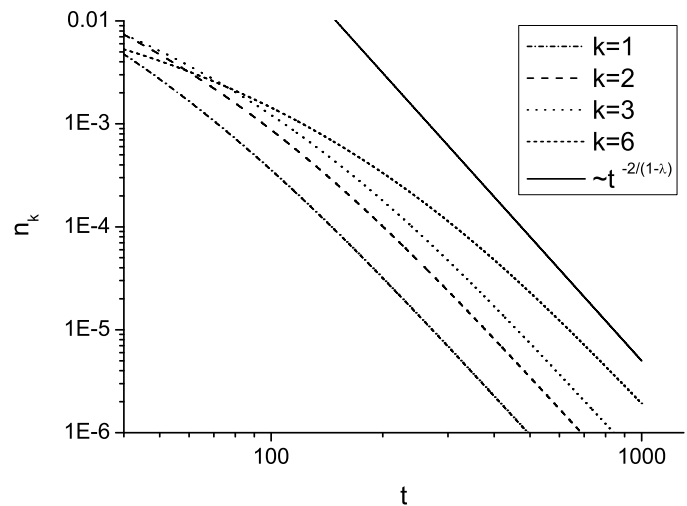}
    }
    \caption
    {The long-time limit behavior of cluster concentrations $n_k$ for the same system
    as in Fig. \ref{C1A05MC}. In accordance with the theoretical predictions,
    Eqs. (\ref{eq:n1t}) and (\ref{eq:n_k}), the cluster concentrations $n_k(t)$ decay
    for $t \gg 1$ with the same slope $t^{-2/(1-\lambda)} \sim t^{-4}$, shown by
    the solid line.
    }
  \label{C1A05version1}
\end{figure}

\begin{figure}[htbp] 
  \centerline{
    \includegraphics[width=0.65\columnwidth]{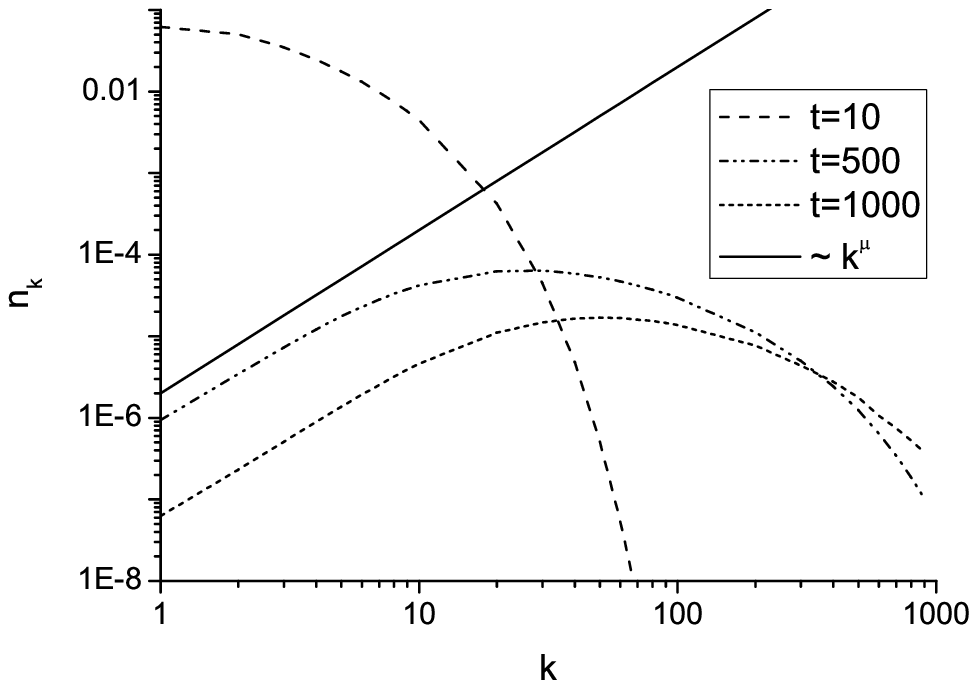}
    }
    \caption
    {
    The cluster mass distribution at different time instants for the same system
    as in Fig. \ref{C1A05MC}. The initial cluster distribution at $t=10$ (long-dashed line)
    drastically differs from that in the scaling regime, $t \gg 1$. The dotted-dashed and dashed
    lines show respectively  the cluster mass distribution for $t=500$ and $t=1000$. The
    solid line shows the theoretical prediction, Eq. (\ref{eq:n_k}), $n_k\sim k^{\mu}$, with
    $\mu=2\lambda/ (1-\lambda)=2$.
    }
  \label{C1A05distrib}
\end{figure}

\begin{figure}[htbp]  
  \centerline{
    \includegraphics[width=0.65\columnwidth]{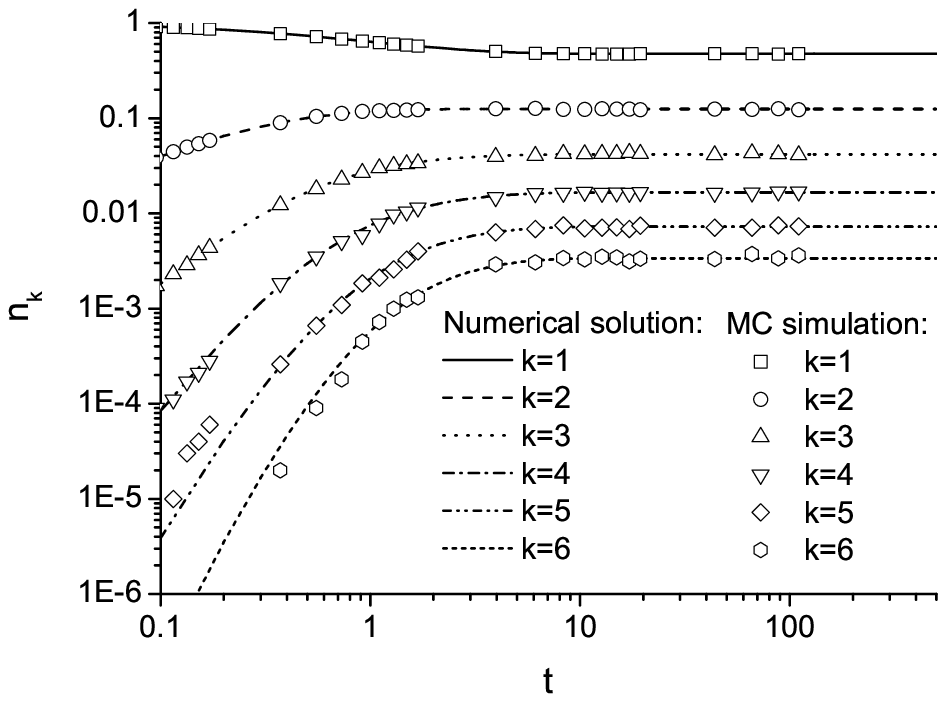}
    }
    \caption
    {
    Evolution of cluster concentrations $n_k(t)$ for  the case of constant
    kinetic coefficients with $\lambda =2$. After a certain period of time the system relaxes to
    a steady state. Lines correspond to the numerical solution of  1000 differential equations,
    symbols -- to the results of MC simulation (100 000 monomers) for the monodisperse initial
    conditions.
    }
  \label{A2C1}
\end{figure}

\begin{figure}[htbp] 
  \centerline{
    \includegraphics[width=0.65\columnwidth]{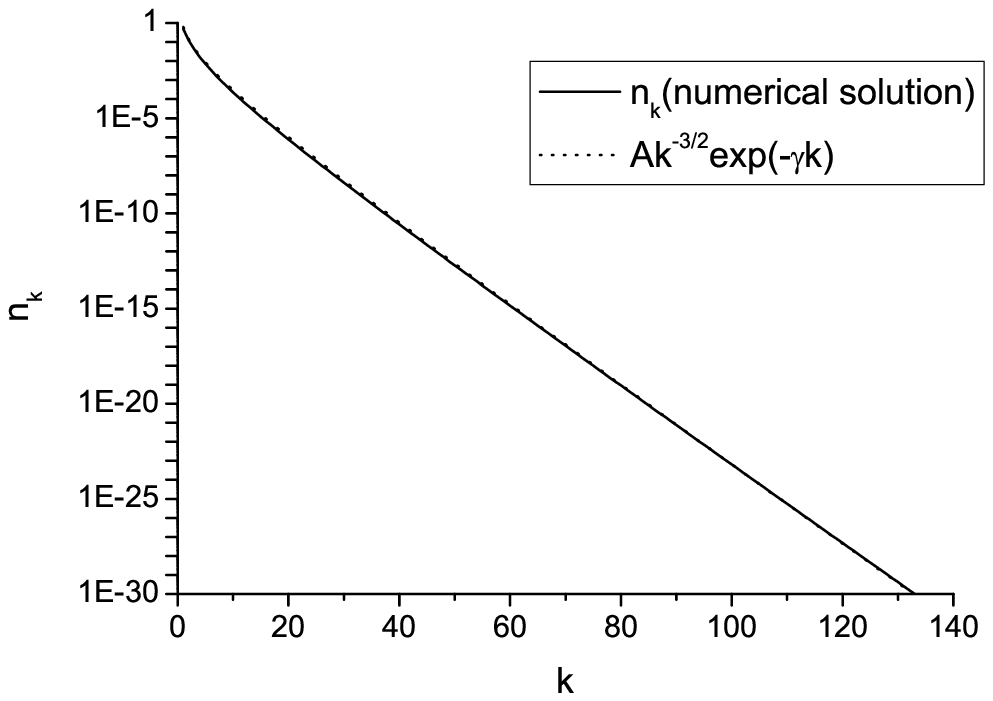}
     }
    \caption
    {
    The steady state distribution of cluster mass for the same system as in Fig. \ref{A2C1}.
    The solid line corresponds to the numerical solution, the dotted line -- to the theoretical
    prediction, Eq. (\ref{eq:nk_asymp}), $n_k = Ak^{-3/2}e^{-\gamma k}$. The constant
    $\gamma= 0.465$, obtained by fitting, is very close to the theoretical value of
    $\gamma= 0.495$, Eq. (\ref{eq:gamma}).
    }
  \label{A2C1distrib}
\end{figure}

\begin{figure}[htbp] 
  \centerline{
    \includegraphics[width=0.65\columnwidth]{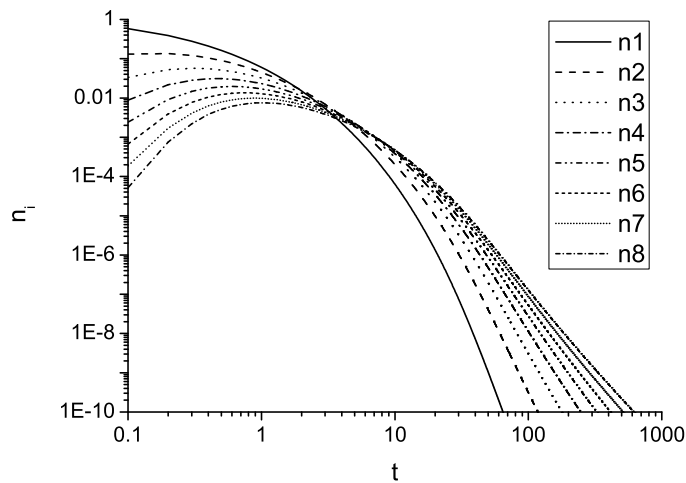}
    }
    \caption
    {
    Evolution of cluster concentrations $n_k(t)$ for the case of ballistic kinetic
    coefficients with constant aggregation and fragmentation energies
    $E_{\rm agg}/T=0.9$, $E_{\rm frag}/T=3$ and $\lambda <1$. The cluster dimension is $D=3$.

    }
  \label{E109E23}
\end{figure}

\begin{figure}[htbp] 
  \centerline{
    \includegraphics[width=0.65\columnwidth]{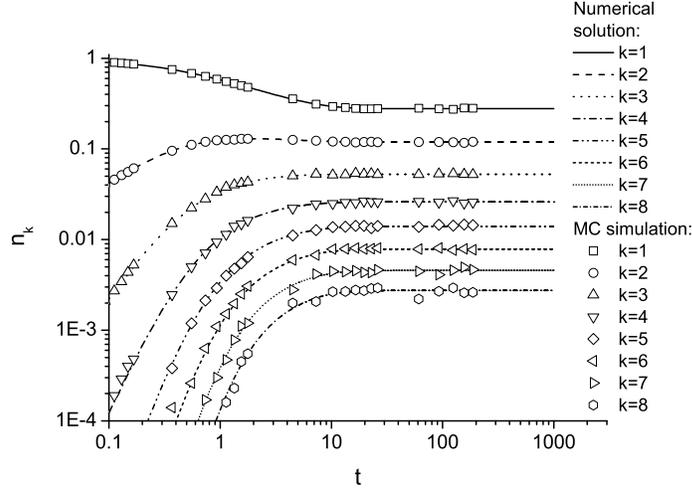}
    }
    \caption{
    Evolution of cluster concentrations $n_k(t)$ in the case of ballistic kinetic coefficients
    with constant aggregation and fragmentation energies $E_{\rm agg}/T=0.3$, $E_{\rm frag}/T=3$ and
    $\lambda >1$. The cluster dimension is $D=3$.
    Similar to the case of constant kinetic coefficients, the system relaxes  to a steady state.
    Lines correspond to the numerical solution of 1000 equations,
    symbols -- to the results of MC simulation (100 000 monomers) for the monodisperse initial
    conditions.
    }
  \label{E103E23}
\end{figure}

\begin{figure}[htbp] 
  \centerline{
    \includegraphics[width=0.65\columnwidth]{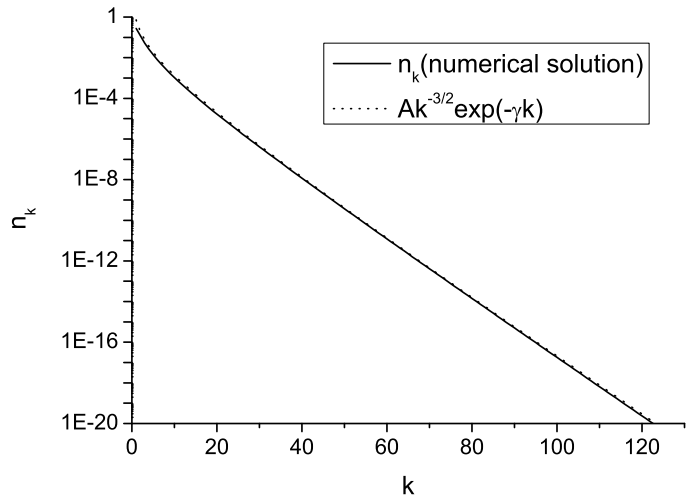}
    }
    \caption
    {
    steady state cluster mass distribution of $n_k$ for the same system as in Fig. \ref{E103E23}.
    Similar to the case of constant kinetic coefficients with $\lambda >1$, the distribution $n_k$
    has a nearly-exponential form, Eq. (\ref{eq:nk_asymp}), $n_k = Ak^{-3/2}e^{-\gamma k}$. The constant
    $\gamma= 0.315$ is obtained by fitting.
    }
  \label{DistribE103E23}
\end{figure}

\begin{figure}[htbp] 
  \centerline{
    \includegraphics[width=0.65\columnwidth]{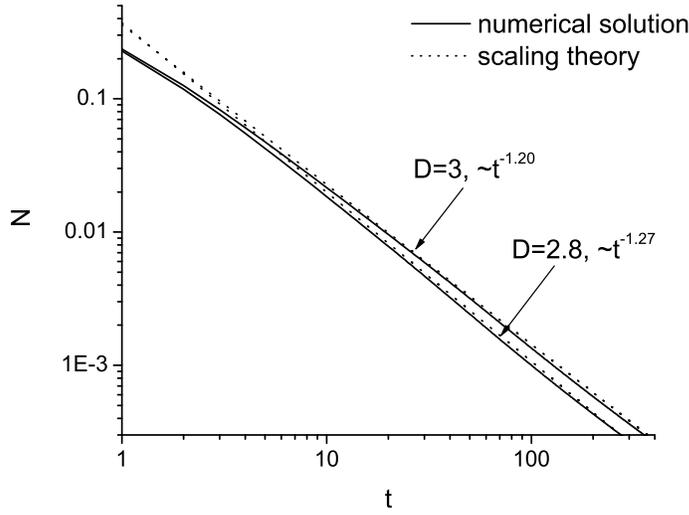}
    }
    \caption
    {
    Evolution of total number of clusters $N(t)$ for  the case of ballistic kinetic
    coefficients with constant aggregation and fragmentation energies
    $E_{\rm agg}/T=0.9$, $E_{\rm frag}/T=3$ and $\lambda <1$ for different cluster dimensions $D$.
    Solid lines correspond to the numerical solution of 1000 differential equations, dotted lines
    show the prediction of the scaling theory,  Eq. (\ref{eq:N_z_sc}).
    }
  \label{Scaling}
\end{figure}

\begin{figure}[htbp] 
  \centerline{
    \includegraphics[width=0.65\columnwidth]{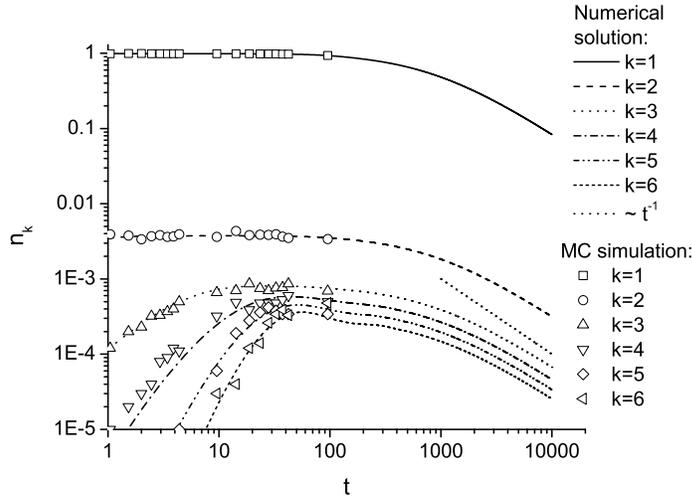}
     }
    \caption
    {
    Evolution of the cluster concentrations $n_k(t)$ for  the case of ballistic coefficients
    $C_{ij}$, $A_{ij}$ with the mass-dependent aggregation and fragmentation energies
    $E_{\rm agg}(i,j)$ and $E_{\rm frag} (i,j)$, given by Eqs. (\ref{eq:E_agg_ij}) and (\ref{eq:E_frag_ij})
    with $E_{\rm agg}^0/T = 0.1$, $E_{\rm frag}^0/T =0.6$. Lines correspond to
    the numerical solution of 1000 differential equations, symbols  -- to the results of MC simulation
    (100 000 monomers) for the monodisperse initial conditions. The cluster dimension is $D=3$.
    Note that the system
    tends initially to a steady state, clearly seen for the first few clusters masses. Later its evolution
    alters to the regime  corresponding to the unlimited cluster growth. In this regime the cluster
    concentrations decay with a slope close to $\sim t^{-1}$, showed in the figure by the dotted line).
    }
  \label{ballistic}
\end{figure}

The results presented on Figs. \ref{C1A05MC}--\ref{ballistic} confirm our theoretical predictions
qualitatively and quantitatively.  For constant rates two opposite types of evolution have  indeed been
observed: the relaxation to a steady state for dominating  fragmentation ($\lambda>1$) and the unlimited
cluster growth when aggregation prevails ($\lambda <1$). Both numerical approaches (the solution of the
differential equations and MC) yield very close results.

In Fig.~\ref{C1A05MC} the evolution of the concentration of clusters of different mass is shown for  the
$\lambda < 1$ regime when the cluster growth continues ad infinitum. Note that all concentrations
$n_k(t)$, except for $n_1(t)$ which always decays, initially increase and then decay to zero.
Figure \ref{C1A05Nkl} shows that the decay of cluster density $N(t)$ well agrees with the  theoretical
prediction (\ref{eq:Nt}).
Figures \ref{C1A05version1}, \ref{C1A05distrib} show respectively the asymptotic evolution of cluster
concentrations and the distribution of the cluster mass for $x=k/t \ll 1$. The  theoretical predictions,
Eqs.~(\ref{eq:n1t}) and (\ref{eq:n_k}), are in a good agreement with the simulations.
Relaxation to a steady state in the case when fragmentation dominates ($\lambda >1)$ is illustrated  in
Fig.~\ref{A2C1}, while Fig.~\ref{A2C1distrib} demonstrates the corresponding stationary cluster mass
distribution.

Note that the numerical simulations confirm the theoretical form of the steady state cluster mass
distribution.
Qualitatively similar behavior is observed for the case of the ballistic kinetic coefficients,
Eqs.~(\ref{eq:KinCoef}), with the constant aggregation and fragmentation energies.
Again, for $\lambda <1$, as for constant kinetic coefficients, clusters unlimitedly grow,
Fig.~\ref{E109E23}, while for $\lambda > 1$ the system relaxes to a steady state, Fig. \ref{E103E23}.
The cluster mass distribution in the steady state may be anew very well fitted with the
nearly-exponential form, Eq.~(\ref{eq:nk_asymp}), see Fig.~\ref{DistribE103E23}.

In Fig.~\ref{Scaling} the prediction (\ref{eq:N_z_sc}) of the scaling theory is compared with  the
numerical data. Again we see that the agreement between the theory and simulations is rather
satisfactory.
Finally Fig.~\ref{ballistic} illustrates evolution of the system with the ballistic coefficients that
depend on the cluster mass in accordance with  Eqs.~(\ref{eq:E_agg_ij}) and (\ref{eq:E_frag_ij}). It is
interesting to note that the system tends initially to a quasi-steady state, as previously for the case
of $\lambda >1$, but then a cross-over to a different evolution regime, corresponding to $\lambda <1$
takes place. In the latter regime all cluster concentrations decay with a similar slope, close to
$t^{-1}$, still to be explained theoretically.


\section{Conclusion}

We analyzed the dynamics of a system where particles move ballistically and undergo collisions which can
lead to decrease or increase of the number of particles. The precise outcome depends on the kinetic
energy $E_{\rm kin}$ in the center-of-mass reference frame. We proposed a simple model with two
threshold energies, $E_{\rm agg}$ and $E_{\rm frag}$, which define a type of an impact: For $E_{\rm
kin}< E_{\rm agg}$ the colliding particles merge, for $E_{\rm agg} \, < \, E_{\rm kin}  \, < \, E_{\rm
frag}$ they rebound, and for $E_{\rm frag} \, < \, E_{\rm kin} $ one the particles (the larger one)
splits upon the collision. We assume that the aggregates are composed of $1, 2, \ldots, k, \ldots$
monomers and split into two equal (for an even number of monomers in the cluster) or almost equal (for
an odd number of monomers) pieces. The monomers are assumed to be stable, that is, they do not further
split. For this model we wrote the Boltzmann kinetic equation for the mass-velocity distribution
function of the aggregates and derived rate equations for the time evolution of the cluster
concentrations. The {\em ballistic} rates were obtained in terms of the aggregation and fragmentation
energy thresholds $E_{\rm agg}$ and $E_{\rm frag}$, masses of the colliding particles and the
temperature of the system (which was assumed to be constant). The Maxwellian velocity distribution for
all species in the system was also assumed.

We analyzed theoretically and studied numerically the rate equations. In the numerical studies we used
two different methods -- the solution of the system of differential equations and Monte Carlo modeling.
Both numerical methods yielded very close results. We started with the simplest case of  constant rates
and observed two opposite evolution regimes  --- the regime of unlimited cluster growth and of the
relaxation to a steady state; we described both these cases analytically. For the regime of the
unlimited cluster growth we obtained the asymptotic time dependence for the cluster concentrations and
for their mass distribution. For the relaxation regime, which corresponds to the prevailing
fragmentation, we derived the asymptotic behavior of the stationary  mass distribution. In the evolving
regime, the cluster concentrations decay as a power law in time;  the stationary mass distribution has a
nearly exponential form. Theoretical  predictions are in a good agreement with numerical results.

We also studied the case of mass-dependent rates arising in the situation when aggregation and
fragmentation energy thresholds are constant. We observed that the  behavior of the system is
qualitatively similar to that of the system with the constant rates.  Surprisingly, we detected that the
steady state cluster mass distribution has also a near-exponential form.   We developed a scaling theory
for the asymptotic large-time behavior of the cluster concentrations and checked  it numerically for
different fractal dimensions of the aggregates. The numerical data agree well with the results of our
theory.

Finally, we explored numerically the case of the ballistic kinetic coefficients with the aggregation and
fragmentation energies depending on the mass of colliding particles. For the aggregation energy
threshold  we use the available in literature result for a collision of particles with surface adhesion.
For the fragmentation energy threshold we adopted a model where $E_{\rm frag}$ is proportional to the
surface energy of the maximal cross-section of the larger particle in the colliding pair. For this model
the dependence on mass of  $E_{\rm frag}$ is much stronger than that of $E_{\rm agg}$. As the result,
the evolution of the system,  where the fragmentation initially prevails and drives it to a steady
state,  alters at later time when the unlimited cluster growth eventually wins and then it continues ad
infinitum.

\end{document}